\documentclass[aps,prb,superscriptaddress,twocolumn,tightenlines]{revtex4}

\usepackage {graphicx}
\usepackage {amsmath}
\usepackage {amsthm}
\usepackage {amsfonts}
\usepackage {amssymb}
\usepackage {mathrsfs}

\usepackage{epstopdf}

\newcommand {\be}{\begin{eqnarray}}
\newcommand {\ee}{\end{eqnarray}}

\begin{document}

\title{Order of Degrees of Freedom in Underdoped Manganites}

\author {E.R.\ S\'anchez Guajardo}
\email {edmundo@lorentz.leidenuniv.nl}
\affiliation {Instituut-Lorentz, 
Universiteit Leiden, P.O.\ Box 9506, 2300 RA Leiden, 
The Netherlands}

\date{\today}

\begin{abstract}
Using the degenerate double exchange Hamiltonian with on-site Coulomb interactions we show how the spin, charge, and orbital state degrees of freedom vary in La$_{1-x}$Ca$_{x}$MnO$_{3}$  for dopings in between $x=4$ and $x=5$.  With the ordered configuration of the system we investigate the band structure for different values of the doping.
\end{abstract}

\maketitle

A large family of Mott insulators are transition metal oxides with a perovskite structure.  Recently, manganites, with their particular colossal magnetoresistance, have been the subject of intense study\cite{jvdbnat, feiner, exp, little, orla, graaf, grenier,coinf}.  Interestingly, these compounds are very rarely both magnetic and ferroelectric, which would allow for technological applications.  It has been shown that charge order results in a lack of inversion symmetry that gives rise to an electric dipole moment\cite{jvdbnat,cohen}.  In the present paper we extend the previous study on the interaction between charge, orbital, and spin ordering in underdoped manganites\cite{jvdbprl,jvdbnat}.

The main complications in finding the charge, orbital, and spin order are the anisotropy of the orbital states per site, and the required dependence of the spin canting on the doping.  The fivefold degeneracy of the $d$ orbital states of the Mn ion is partially lifted by the crystal field generated by the surrounding oxygen atoms in the perovskite structure.  The result is twofold degenerate $e_{g}$ orbital states, d$_{x^{2}-y^{2}}$ and d$_{3z^{2}-r^{2}}$, and threefold degenerate $t_{2g}$ orbital states, d$_{xy}$, d$_{yz}$, and d$_{zx}$.  The degeneracy of the $e_{g}$ states is further lifted by the lattice distortions, or by the on-site Coulomb interactions. The difference is that the lattice distortions lift the degeneracy completely by lower the symmetry of the cubic crystal which lowers the energy of the  d$_{3z^{2}-r^{2}}$ state while raising that of  d$_{x^{2}-y^{2}}$.  The on-site Coulomb interactions make the system quasidegenerate because the degeneracy is only lifted after either of the two orbital states is first occupied.  In an array of manganese ions the orbital states in neighboring ions may be rotated with respect to each other.  After choosing a basis we may characterize the orbital states of every ion by an angle\cite{feiner}.  

The electronic configuration of each ion in the array may also vary.  The Mn$^{3+}$ ion in the parent compound of interest, LaMnO$_{3}$, has an open shell configuration in which three electrons are in the $t_{2g}$ orbitals and one in either of the two $e_{g}$ orbitals.   The $t_{2g}$ electrons are strongly stabilized by the crystal field, and have a lower hybridization with the 2p states of the oxygen atoms, which results in the electrons being localized\cite{tokura}.  When doping with a divalent ion that introduces holes, for instance La$_{1-x}$Ca$_{x}$MnO$_{3}$, some sites change to a Mn$^{4+}$ closed shell configuration in which the $e_{g}$ orbitals are empty.

There are three interactions that reproduce the properties of the solids in discussion:  the double-exchange, the superexchange, and the on-site Coulomb interaction.  The double-exchange has an energy scale given by the hopping amplitude $t$.  We consider $t=0.622$ eV, which is found in the literature\cite{millis}.  It leads to the formation of bands and to the conduction of electrons.  The superexchange is a magnetic interaction between localized spins of neighboring sites with an energy scale $J$.  The on-site Coulomb interaction is the electrostatic interaction between electrons in different $e_{g}$ orbital states in the same Mn ion.  Its energy scale is  $U$.  In the Mott insulator the otherwise itinerant electrons are localized due to $U>>t$.

\begin{figure} [tb]
\includegraphics[width=.4\textwidth]{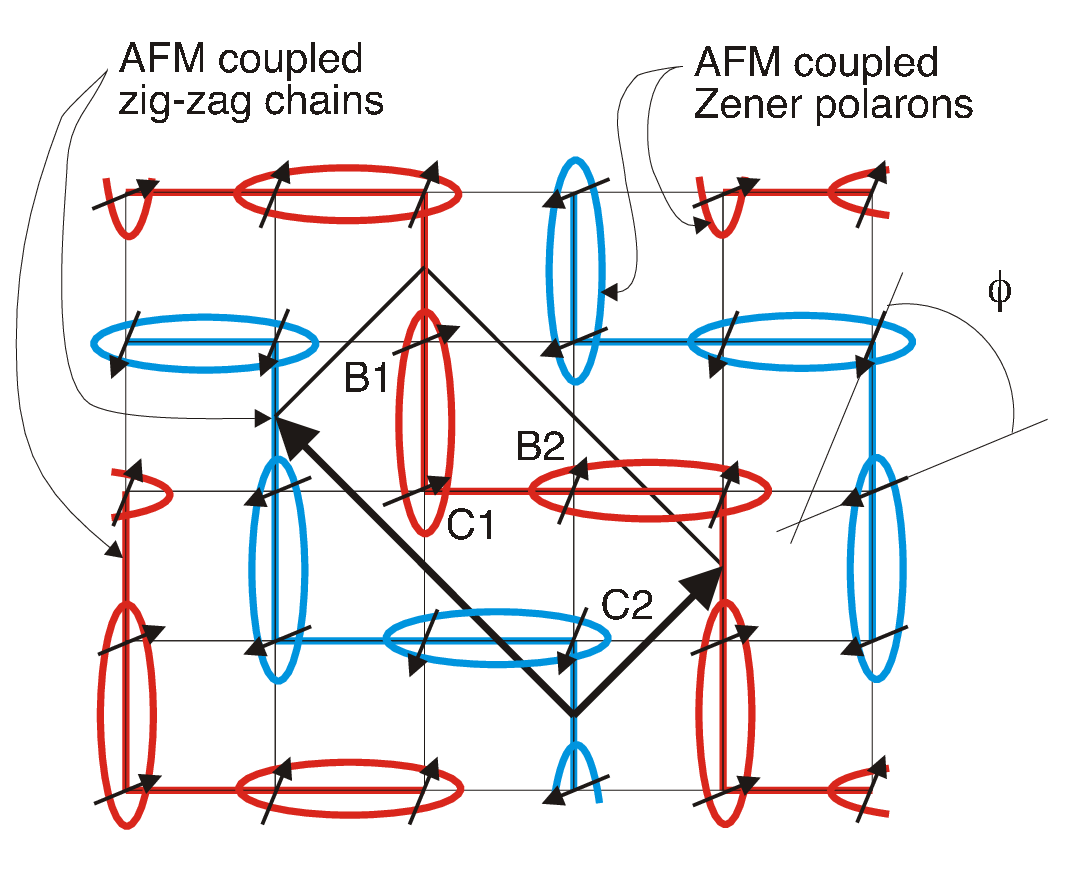}
\caption{\label{unit}The lattice with the corresponding unit cell for the Zener polaron phase phase $\phi=\pi/2$ (top), the intermediate phase $\phi=\pi/4$ (middle), and the CE phase $\phi=0$ (bottom).  The rotation of the spins is depicted for each spin canting $\phi$.  The Zener polaron phase shows dimers, with blue and red dimers AFM coupled within a diagonal.  The intermediate phase shows the superposition of the Zener polaron and the CE phase.  The CE phase is composed of AFM coupled zig-zag chains.\cite{ersg}}
\end{figure}

We consider two limiting cases of La$_{1-x}$Ca$_{x}$MnO$_{3}$:  the Zener polaron phase at $x=0.4$ doping, and the CE phase at $x=0.5$.  Both phases are known to be insulating.  As shown in Fig.~\ref{unit}, the unit cell is composed of four sites, with one Mn ion each.  The CE phase is made up of zig-zag chains that have ferromagnetic spin order within the chain, and that are antiferromagnetically coupled with the neighboring chains.  This phase presents the checkerboard charge order in which sites alternate between Mn$^{3+}$ and Mn$^{4+}$.  The orbital order is d$_{3z^{2}-r^{2}}$ and d$_{x^{2}-y^{2}}$ in the corner sites (C1 and C2), d$_{3x^{2}-r^{2}}$ in the horizontal bridge (B2), and d$_{3y^{2}-r^{2}}$ in the vertical bridge (B1).  The Zener polaron phase is made up of dimers with equivalent sites: B1 and C1, and B2 and C2 become equivalent.  The spins within a dimer are parallel, while spins in between neighboring dimers with the same crystallographic direction couple antiferromagnetically.  Perpendicular dimers have perpendicular spin orientations with respect to each other.  The charge in a dimer is delocalized in between the equivalent sites, which results in a bond-centered charge order~\cite{jvdbnat}.  

The Hamiltonian of the system is given by
\begin{eqnarray}
H&=&t\sum_{<ij>,\sigma\sigma^{\prime}} \left<\chi_{i}|\chi_{j}\right>c_{i,\sigma}^{\dagger}c_{j,\sigma^{\prime}}\Gamma_{ij}^{\sigma\sigma^{\prime}} \nonumber\\
& &{  }+J_{AF}\sum_{<ij>}\vec{S}_{i}\cdot\vec{S}_{j} + U\sum_{i,\sigma\neq\sigma^{\prime}} n_{i}^{\sigma}n_{i}^{\sigma^{\prime}},
\label{Hamiltonian}
\end{eqnarray}
where $i,j$ are site labels, and $\sigma,\sigma^{\prime}$ denote $e_{g}$ orbital states. The sum  $<ij>$ is nearest neighbor sites.  The densities $n_{i}^{\sigma}$ are per orbital per site, and $\vec{S}_{i}$ are the localized (classical) $t_{2g}$ spins in nearest neighbor sites.  We consider an antiferromagnetic superexchange interaction, $J_{AF}$.  The kinetic term contains the fermionic creation and annihilation operators, $c_{i,\sigma}^{\dagger}$ and $c_{j,\sigma^{\prime}}$, in between orbitals of neighboring sites, and the symmetry matrix element $\Gamma^{\sigma\sigma^{\prime}}_{ij}$ that takes into account the overlap of the orbitals involved and the crystallographic direction of the hopping.  We consider the overlap of the spin wave functions\cite{nagaosa}, $\left<\chi_{i}|\chi_{j}\right>$, for the hopping between nearest neighbor sites.  It depends on the relative direction between the two spins, $|\left<\chi_{i}|\chi_{j}\right>|=\cos(\phi_{ij}/2)$, where $\phi_{ij}$ is the relative angle between two spins.  The hoping is largest for parallel spins, while antiparallel spins cannot hop.

We consider the zero temperature limit with strong ferromagnetic Hund's coupling between localized $t_{2g}$ and itinerant $e_{g}$ electrons, $J_{H}/t\rightarrow\infty$.  The antiferromagnetic interaction becomes  $J_{AF}\sum_{<ij>}|S|^{2}\cos{\phi_{ij}}$.  With respect to the symmetry matrix, we choose our orbital basis states to be $\alpha=3z^{2}-r^{2}$, and $\beta=x^{2}-y^{2}$.  In that case we have
\begin{eqnarray}
\Gamma_{<ij>//y}=\frac{1}{4}\left[
\begin{array} {cc}
1 &\sqrt{3}\\
\sqrt{3} &3
\end{array}
\right] ,
\label{gamma}
\end{eqnarray}
that takes into account the overlaps of the $e_{g}$ orbitals involved when hoping along the $y$-axis\cite{dagotto}.  The hopping along the $x$-axis has a phase with respect to the $y$-axis which results in $\Gamma^{\alpha\beta}_{<ij>//x}=\Gamma^{\beta\alpha}_{<ij>//x}=-\sqrt{3}$.  After choosing a basis we may refer the orbital states of every ion, $\alpha_{i}$ and $\beta_{i}$, to the reference basis with the transformation
\begin{eqnarray}
\left[
\begin{array} c
|\alpha_{i}(\xi)>\\
|\beta_{i} (\xi)>
\end{array}
\right]&=& \left[
\begin{array} c
\cos(\frac{\xi}{2})|\alpha > + \sin(\frac{\xi}{2})|\beta >\\
-\sin(\frac{\xi}{2})|\alpha> + \cos(\frac{\xi}{2})|\beta>
\end{array}
\right].
\label{rotation}
\end{eqnarray}
Finally, we take the on-site Coulomb interaction to be $U/t=4$ and use the mean-field approximation of the densities $<n_{i}^{\sigma}>_{mf}=<c_{i,\sigma}^{\dagger}c_{i,\sigma}>$.

After decoupling the on-site Coulomb interaction, $n_{i}^{\alpha}n_{i}^{\beta}\approx <n_{i}^{\alpha}>n_{i}^{\beta}+n_{i}^{\alpha}<n_{i}^{\beta}>-<n_{i}^{\alpha}>< n_{i}^{\beta} >$, we first calculate the charge order (CO), orbital order (OO), and magnetic order (SO) for different fixed values of the doping, varying the spin canting angle from $\phi_{ij}=0$ (CE phase) to $\phi_{ij}=\pi/2$ (Zener polarons).  Taking the relative angle between spins to change equally we drop the site indices and in what follows use $\phi$.  We also bear in mind that the magnetic correlation between some spins does not change.  The system is taken into account by rotating the orbital states per site to our working bases, $\alpha$ and $\beta$, with the transformation
\begin{eqnarray}
n_{i}^{\alpha}&=&n_{i}^{\alpha_{i}}\sin^{2}(\xi_{i}/2)+n_{i}^{\beta_{i}}\cos^{2}(\xi_{i}/2)\nonumber\\
n_{i}^{\beta}&=&n_{i}^{\alpha_{i}}\cos^{2}(\xi_{i}/2)+n_{i}^{\beta_{i}}\sin^{2}(\xi_{i}/2)
\label{rotation}
\end{eqnarray}
before solving numerically.   The self consistency equations,
\begin{eqnarray}
\xi_{i}&=&\tan^{-1}(2<c_{i,\alpha}^{\dagger}c_{i,\beta}>/(\tilde{n}_{i}^{\alpha}-\tilde{n}_{i}^{\beta}))\\
n_{i}^{\alpha_{i}}&=&\tilde{n}_{i}^{\alpha}\cos^{2}(\xi_{i}/2)+<c_{i,\alpha}^{\dagger}c_{i,\beta}>\sin(\xi_{i})\nonumber\\
&&{  }+\tilde{n}_{i}^{\beta}\sin^{2}(\xi_{i}/2)\nonumber\\
n_{i}^{\beta_{i}}&=&\tilde{n}_{i}^{\alpha}\sin^{2}(\xi_{i}/2)-<c_{i,\alpha}^{\dagger}c_{i,\beta}>\sin(\xi_{i})\nonumber\\
& &{  }+\tilde{n}_{i}^{\beta}\cos^{2}(\xi_{i}/2),\nonumber
\label{selfconsistent}
\end{eqnarray}
enforce that the rotation of the calculated densities, $\tilde{n}_{i}^{\sigma}$, from the working basis back to the on-site basis results in a diagonal on-site Coulomb interaction.  

\begin{figure} [tb]
\includegraphics[width=0.36\textwidth]{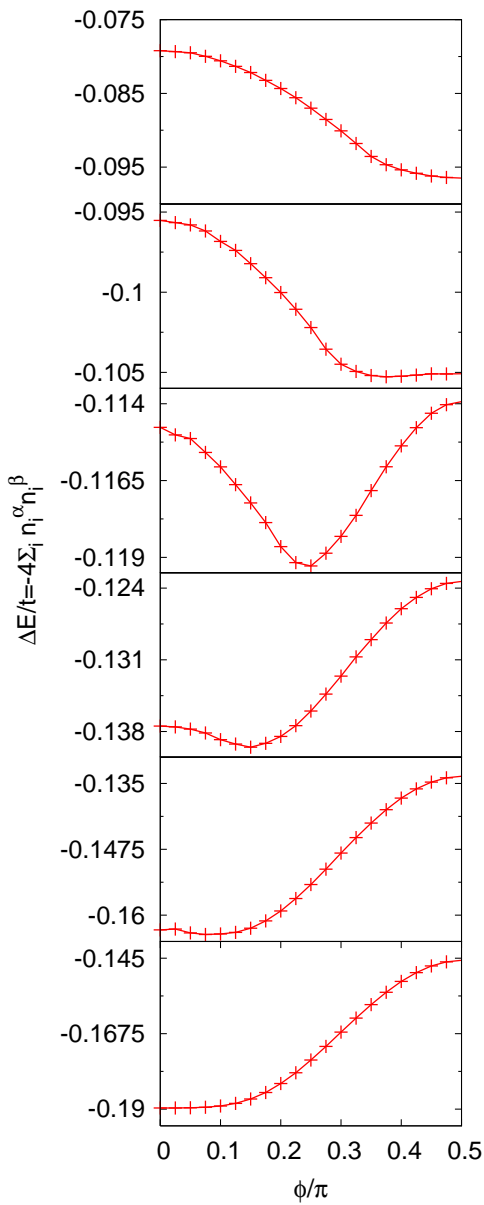}
\caption{\label{minimo}Contribution to the energy from the quadratic term in the mean-field value of the densities for fixed doping as a function of spin canting for $U/t=4$.  For different doping values between the Zener polaron phase (top) to the CE phase (bottom) with $\Delta x=0.02$ increments.  Notice the change of the minimum value for each doping as a function of $\phi$.}
\end{figure}

Next, we calculate the total energy of the system and identify the ground state energy, $E_{0}(\phi)$, for fixed doping.  The total energy is composed of a term linear in the mean-field value of the densities, and a term quadratic in the mean-field value of the densities.  The linear term, namely, the kinetic term, increases monotonously when changing the spin canting from one limiting value of the doping to another.  On the other hand, the quadratic term has a clear minimum that moves from one limiting doping to the other as the spin canting changes, shown in Fig.~\ref{minimo}.  It is this term that, when added to the linear term, gives a minimum to the total energy.  Hence, the on-site Coulomb interactions stabilize the system.  We found that for larger ratios of $U/t$ the range of stability goes below 0.4 doping, but the validity of the numerical calculation is lost as we go beyond the ratio for which mean-field is valid. 

Following the determination of $E_{0}(\phi)$ for different doping, we stablish the relation between the spin canting $\phi$ and the doping $x$ by fitting the curve $\phi(x)$ given by the energy minima.  We use $\phi(x)$  to find the ground state energy, $E_{0}[\phi(x)]$, as the doping $x$ changes continuously from 0.4 to 0.5, {\it i.e.}, the stable configurations of charge, orbit, and spin order. 

With respect to our results for the orbital order, we first need to understand the difference between corner and bridge orbital states for the CE phase, $\xi=\pm\pi/3$.  We bear in mind\cite{feiner} that pairs of orthogonal orbital states differ by an angle of $\pi$, $|\xi>$ and $|\xi+\pi>$.  Since our basis orbital states are chosen to be $|0>=$3z$^{2}$-r$^{2}$ and $|\pi>=$x$^{2}$-y$^{2}$, and the orbital states at the bridge sites having nonzero occupation are $|\pi/3>=$3y$^{2}$-r$^{2}$ and $|-\pi/3>=$3x$^{2}$-r$^{2}$, we conclude that the reference orbital state for $\xi=\pm\pi/3$ is $|\pi>=$x$^{2}$-y$^{2}$.  This allows us, on the one hand, to interpret the found mean field densities of orbital states per site by noticing that the labels $\alpha$ and $\beta$ at the bridge sites are interchanged with respect to the corner sites.  That is to say, we have to take into account the phase difference in orbital states when the charge moves from a $|\xi+\pi>$ orbital state to a $|\xi>$ one.  On the other hand, we can expand on the idea that the on-site Coulomb interaction pushes the charge from the corner into the bridge sites\cite{jvdbprl}.  Previously, no reference to the specific orbitals involved in the hopping was made.  We conclude that the hopping takes place from the x$^{2}$-y$^{2}$ orbital states at the corner sites unto the 3y$^{2}$-r$^{2}$ and 3x$^{2}$-r$^{2}$ sites at the bridges given the required rotation of orbital states.  Since the difference is exactly $|\pi/3|$ it seems to suggest that all the hopping between bridge and corner sites take place in the above mentioned orbitals, and none from the 3z$^{2}$-r$^{2}$ orbital state at the corners into the bridges.  This is in agreement with the findings of other computational schemes\cite{grenier}.

\begin{figure} [tb]
\includegraphics[width=0.39\textwidth]{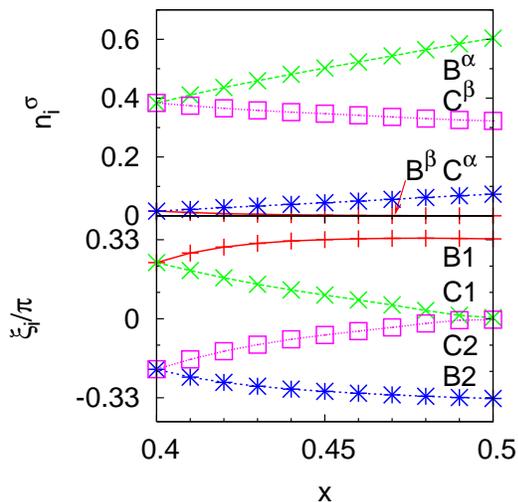}
\caption{\label{resultados}The charge order (top) and orbital order (bottom) for different doping.}
\end{figure}

Interestingly, unlike the change of density per orbital state, the angle $\xi_{i}$ does not change linearly with the doping for both bridge and corner sites orbital states.  As seen in Fig.~\ref{resultados}, the orbital states at the bridge sites remain almost constant between $x=0.45$ and $x=0.5$.  For the Zener polaron phase we find that the orbital states per site for each dimer is the same, as required for the charge to be delocalized and centered on the bond\cite{jvdbnat}.  In this limit we find that the orbital states are defined by $\xi=\pm\pi/4$, with the angular difference between the orbital state in dimers along the $y$-axis with respect to those in the $x$-axis equal to $|\pi/2|$.  Thus, the orbital states between horizontal and vertical dimers are orthogonal with respect to each other.  The result is two sublattices of dimers in diagonal stripes that have orthogonal spin and orbital order with respect to each other.    

\begin{figure} [tb]
\includegraphics[width=0.45\textwidth]{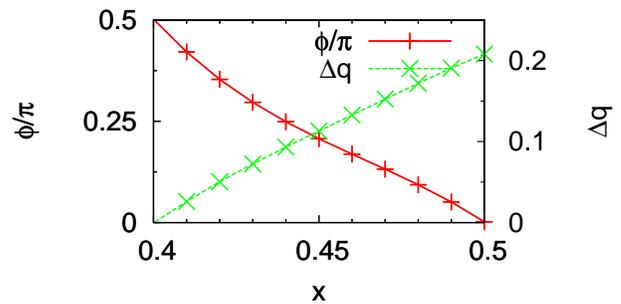}
\caption{\label{chargedisp}Charge disproportion $\Delta q$ and spin canting $\phi$ as a function of doping.  The CE phase shows the largest disproportion between sites, corresponding to the checkerboard charge order, while the Zener polaron has $\Delta q =0$ corresponding to charge ordered on the bond between bridge and corner sites.  For the spin canting we see a continuous rotation between CE and Zener polaron phase.}
\end{figure}

Concerning the charge order, our results reproduce the checkerboard charge order of the CE phase, and the predicted charge order for the Zener polaron phase, which is on the bond\cite{jvdbnat}.  This can be seen in Fig.~\ref{chargedisp}, where the charge disproportionation between two neighboring sites is zero for the Zener polaron phase, and maximum for the CE phase.  Previously, a similar curve had been proposed for $U/t=1$, were the largest valence difference was less than $\Delta q =0.1e$ at the CE phase~\cite{jvdbnat}.  We find for $U/t=4$ a charge disproportionation slightly greater than $\Delta q =0.2e$.  As already known, a stronger on-site Coulomb interaction pushes more charge from the correlated corner sites unto the uncorrelated bridge sites of the CE phase.  This result is far from the one using exact diagonalization but well within the mean-field approximation\cite{jvdbprl}.  We also present in this graph our results for the dependence of the spin canting $\phi$ on the doping.  

The density of particles per orbital state per site is plotted in Fig.~\ref{resultados}.  At $x=0.5$ the corner sites have more charge in the orbital state d$_{x^{2}-y^{2}}$ (C$^{\beta}$), than in the d$_{3z^{2}-r^{2}}$ (C$^{\alpha}$).  On the bridge sites all the charge is in the orbital states d$_{3x^{2}-r^{2}}$/d$_{3y^{2}-r^{2}}$ (B$^{\beta}$), while the B$^{\alpha}$ orbital is empty.  Hence, the bridge sites are uncorrelated, as already shown~\cite{jvdbprl}.  Notice how the aforementioned exchange of labels for the orbital states is crucial in interpreting the densities properly.  The densities change linearly as a function of doping.  As the doping decreases from the CE phase we see two main effects:  the densities C$^{\alpha}$ and B$^{\beta}$ become equal as well as C$^{\beta}$ and B$^{\alpha}$, and the density B$^{\beta}$ becomes nonzero.  Hence, the corner and bridge sites at  $x=0.4$ become equivalent, resulting in Zener polaron dimers.  The equivalence between sites results in a delocalization of the charge in between the two, that leads to the charge being centered on the bond on average, as previously mentioned.

\begin{figure} [tb]
\includegraphics[width=0.495\textwidth]{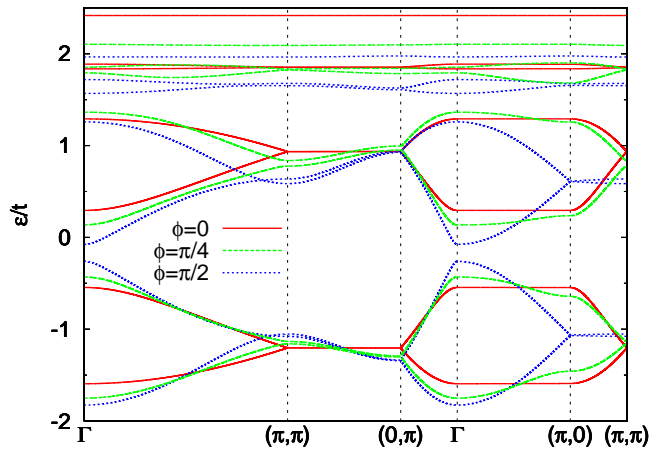}
\caption{\label{labanda}Bands between points of high symmetry of the reciprocal space lattice for $U/t=4$.  Notice the changes of the energy gap between valence and conduction bands, and the closely packed and almost dispersionless bands above the conduction bands.}
\end{figure}

\begin{figure} [tb]
\includegraphics[width=0.49\textwidth]{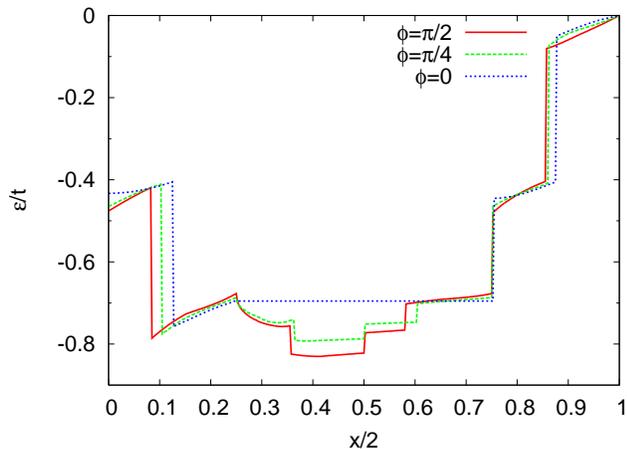}
\caption{\label{energy}Total energy for the limiting case $U\rightarrow 0$.  Notice how the possibility to fill zero energy states changes as soon as the spin canting becomes nonzero.}
\end{figure}

After finding the stable configuration of the system we calculate the band structure for different spin canting values, shown in Fig.~\ref{labanda}.  Doing basic band structure analysis\cite{roessler} we see for the CE phase that the dispersionless bands along $\Gamma-(\pi,0)$ and $(\pi,\pi)-(0,\pi)$ in the reciprocal lattice indeed show antiferromagnetically coupled zig-zag chains.  Moreover, the bands are degenerate along this direction.  We also see how the energy gap at the $\Gamma$ point between the valence bands and the empty conduction bands is the largest with respect to the other spin canting bands shown.  For half-filling all the valence bands are filled and the system is an insulator.  In addition, we also notice that four of the bands are above the conduction bands:  the strong correlation between on-site orbitals for $U\neq0$ makes four bands energetically inaccessible by shifting them above the conduction bands.  This makes the system effectively monovalent.   In the limit $U/t\rightarrow 0$ these bands become fourfold degenerate zero energy bands.  It was previously argued that while in the CE phase one could dope these nondispersive bands causing no change in energy\cite{jvdbprl}, thus possibly occupying both orbital states per site. We add to it that, as seen in Fig.~\ref{energy}, this is clearly not the case as soon as $\phi\neq0$.  For spin canting $\phi=\pi/4$, we see that the bands along $\Gamma-(\pi,0)$ and $(\pi,\pi)-(0,\pi)$ are now active, and cross the Fermi energy $\epsilon^{\phi=\pi/4}_{F}/t=-0.649$ along the segment $\Gamma-(\pi,\pi)$.  Further, their degeneracy has been lifted by the opening of a small energy gap at the boundary of the Brillouin zone.  On the other hand, the energy gap at the $\Gamma$ point between the valence and conduction bands closes.  The energy gap splitting occupied and empty bands in the CE phase in the limiting case $U\rightarrow 0$ was found\cite{jvdbprl}  to be  $\Delta^{CE}_{U\rightarrow 0}/t = 1$.  For $U\neq 0$ the gap reduces to $\Delta^{CE}/t = 0.83$, which is $\Delta^{CE} = 0.51$ eV.  That the energy gap may be small enough to allow for double-exchange has been discussed experimentally\cite{coinf}.  Finally, for the Zener polaron phase at $\phi=\pi/2$ we see that the system is practically dispersionless along the boundaries of the reciprocal lattice $(\pi,0)-(\pi,\pi)$.  Also interesting is that the bands now cross each other along the path $\Gamma-(\pi,\pi)$, and cross the fermi energy $\epsilon^{\phi=\pi/2}_{F}/t=-0.835$.  The energy gap further closes to $\Delta^{ZP}/t=0.57$, which is $\Delta^{ZP}=0.35$ eV.

\begin{figure} [tb]
\includegraphics[width=0.47\textwidth]{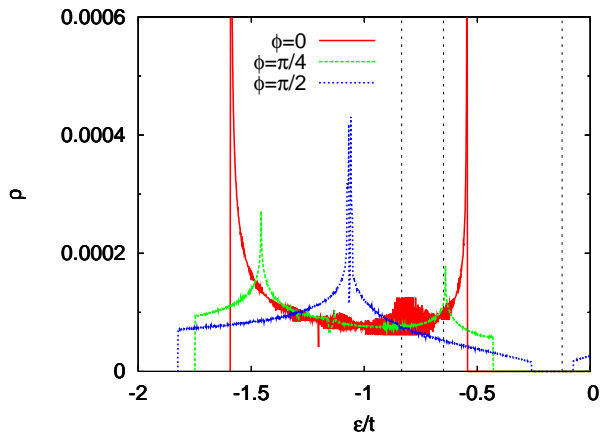}
\caption{\label{lasdosred}Density of states corresponding to the valence bands.  The dotted lines correspond to the values of the Fermi energy of the insulating CE phase (right), intermediate spin canting $\phi=\pi/4$ (middle), and Zener polaron phase (left).}
\end{figure}

Finally, the density of states, $\rho(\epsilon)$, is calculated.  It has sharp narrow peaks corresponding to the bands above the conduction bands.  Focusing on the density of states of the valence electrons, Fig.~\ref{lasdosred}, we notice that the Fermi energy of the CE phase is delocalized at the middle of the gap between valence and conduction states, giving rise to an insulating state.  Interestingly, for the intermediate phase at $\phi=\pi/4$ the Fermi energy coincides with a van Hove singularity.  This is due to the point $(\pi,0)$ of the reciprocal lattice, where the group velocity vanishes. 

We conclude that in underdoped La$_{1-x}$Ca$_{x}$MnO$_{3}$ for doping in between $x=0.4$ and $x=0.5$ the effect of the on-site Coulomb interactions is to make the solid effectively monovalent, and to stabilize the orbital, charge and spin configuration.  We have also shown how the orbital, charge and spin order change when going from the Zener polaron phase to the CE phase.  Regarding the band structure, it was proven that the energy gap closes as the spin canting changes from zero to $\pi/2$, and is smaller than the hopping amplitude, possibly allowing for double-exchange processes.  It has been proposed elsewhere\cite{little} that the inclusion of the oxygens in the calculations results in the CE phase being made up of oxygen stripes.  Since we are considering atomic orbital wave functions our description of charge order in the bond is an assumption that does not allow us to compare our result with this theoretical work.  Our results show that the magnetic correlation between spins within a dimer is always ferromagnetic and does not change for the dopings being discussed.  We suggest that within the limitations of our model both results are related.  We do not find any agreement with the results in which a different configuration of the CE phase was proposed~\cite{orla}.  Future work should consider thermal effects.

The author is deeply grateful to Frank Kr\"uger for lengthy discussions, and to Jens Bardarson for support while writing the code for the numerics.

This work was supported by the Stichting voor Fundamenteel Onderzoek der Materie (FOM), which is supported by the Nederlandse Organisatie voor Wetenschappelijk Onderzoek (NWO).

\end{document}